\documentclass[12pt]{article}
\usepackage[letterpaper]{geometry}
\usepackage[english]{babel}
\usepackage[utf8]{inputenc}
\usepackage{amsmath}
\usepackage{ragged2e}
\usepackage{amssymb}
\usepackage[flushleft]{threeparttable}
\usepackage{graphicx}
\usepackage{tabularx}
\usepackage{footnote}
\usepackage{tablefootnote}
\makesavenoteenv{tabular}
\usepackage{longtable}
\usepackage{setspace}   
\usepackage{subfig}
\IfFileExists{kpfonts.sty}{\usepackage{kpfonts}}{} 
\usepackage{microtype}
\usepackage{booktabs}   
\usepackage{bm}         
\usepackage{listings}   
\usepackage{verbatim}   
\usepackage{color}
\usepackage[colorlinks=true]{hyperref}
\usepackage[colorinlistoftodos]{todonotes}
\usepackage{natbib}

\def\name{Brice Romuald Gueyap Kounga}

\hypersetup{
        urlcolor = blue,
        pdfauthor = {\name},
        pdfkeywords = {Trust, Traditional leaders, Colonization, British, French, Cameroon},
        pdftitle = {The Long-Term Effects of British and French Colonization in Africa on Trust in Traditional Leaders},
        pdfsubject = {},
        pdfpagemode = UseNone,
        citecolor = blue
}

\title{The Long-Term Effects of British and French Colonization in Africa on Trust in Traditional Leaders\footnote{Special thanks to Prof.\ James Tybout and Francis Touola for their comments and suggestions in the preparation of this paper. The views expressed herein and any remaining errors are mine.}
}

\author{
    Brice Romuald Gueyap Kounga\\
    Western University\\
     bgueyapk@uwo.ca
	}
\date{\today}

\begin{document}
\maketitle
\onehalfspacing
\begin{abstract}
\noindent Trust in local institutions matters for trade, public goods provision, conflict resolution, and democratic consolidation. Using individual data from rounds 6 and 7 of the Afrobarometer surveys, I document that respondents in former British colonies are substantially more likely to trust traditional leaders than respondents in former French colonies. Cross-country comparisons of this kind are confounded by unobserved heterogeneity in pre- and post-colonial histories. To make progress on identification, I focus on Cameroon, which contains regions colonized by Britain and regions colonized by France within a single modern state, and I exploit the former colonial partition line as a geographic discontinuity, comparing respondents who live close to either side of the anglophone-francophone boundary. The theoretical argument is that the two colonial powers differed systematically in how they treated customary authority: British indirect rule preserved the governance functions of chiefs, while French direct rule subordinated chiefs to the administration and assigned them its most coercive tasks, taxation and forced labor above all, and these divergent experiences shaped the legitimacy of chieftaincy in ways that persist through institutional continuity and socialization. Consistent with this argument, respondents on the formerly British side are 22 to 26 percentage points more likely to trust traditional leaders than their neighbors on the formerly French side, and they are far more likely to have contacted a traditional leader in the previous year.
\end{abstract}


\section{Introduction}~

There has been growing interest in the role of trust in African economic development, where low levels of trust have been cited as an obstacle to growth \citep{Collier_Gunnin1999(a)}. Trust is a fundamental element of social capital: it sustains well-being, facilitates coordinated action, and improves the efficiency of society \citep[pp.~167]{Robert_Putnam(1993)}. A substantial empirical literature confirms that social trust is an important determinant of political and economic outcomes \citep{knack2002, uslaner_2002, Christian_2006}, and that high-trust societies exhibit lower transaction costs, better-functioning institutions, and faster growth \citep{Knack_and_Keefer(1997), Zak&Knack(2001), Yann_and_Pierre(2010)}. Given that trust is central to economic and political development, it is important to understand where it comes from. A prominent view holds that trust originates in shared values that are shaped by historical and cultural heritage \citep{Putnam2000}, and individual-level evidence confirms that historical experiences leave durable marks on trust \citep{Ferrara, Nunn_Wantchekon_(2011)}.

This paper studies a specific and understudied dimension of trust in Africa: trust in traditional leaders, that is, chiefs and customary authorities. The research question is whether this trust carries a legacy of the form of European colonial rule. Both choices of variable deserve justification. Trust in traditional leaders, rather than interpersonal trust or trust in state institutions, is the appropriate outcome here for two reasons. Substantively, chiefs remain the level of authority with which much of the African population actually interacts: they allocate land, resolve disputes, and mediate access to government, especially in rural areas, and Afrobarometer data show that popular engagement with them rivals or exceeds engagement with elected local officials \citep{Logan2013, Baldwin2016}. Trust in traditional leaders therefore measures the perceived legitimacy of the institutions that govern daily life for millions of citizens, and it remains far less studied than the interpersonal and state-institutional trust examined in the existing literature. Analytically, it is the outcome for which colonial rule offers the sharpest prediction, which is also the answer to why the colonizer is the explanatory variable of interest: Britain and France were the two largest colonial powers in Africa, and the historiography has long emphasized that what most distinguished their administrative strategies was precisely their treatment of customary authority. British administration relied comparatively more on indirect rule through existing native authorities, while French administration was comparatively more centralized, treated chiefs as subordinate agents of the colonial state, and relied more heavily and for longer on coerced labor \citep{Crowder1964, grier1999colonial, Cooper1996, Mamdani1996}. If any historical variable should have left a mark on the citizen-chief relationship, it is this one. Section \ref{sect_theory} develops the argument into explicit channels and predictions.

Two framing points should be fixed at the outset. First, nothing in this paper implies that colonization benefited the colonized. The comparison is strictly relative: colonial rule may well have damaged trust everywhere, and the findings are consistent with British rule simply being less corrosive of chiefly legitimacy than French rule on this one dimension. Second, higher trust in traditional leaders is not treated here as intrinsically good. Some skepticism toward elites is healthy in any polity, and as \citet{Mamdani1996} and \citet{AcemogluReedRobinson2014} emphasize, deference to chiefs can reflect entrenched and unaccountable authority rather than earned legitimacy. The estimates measure the state of the citizen-chief relationship, not its welfare value.

Using individual data from rounds 6 and 7 of the Afrobarometer surveys, I first document a large cross-country gap: 81 percent of respondents in former British colonies report trusting traditional leaders somewhat or a lot, against 64 percent in former French colonies. Such a comparison cannot, by itself, be given a causal interpretation, because colonizer identity is bundled with everything else that differs across modern states. I therefore pursue two strategies that progressively tighten identification. First, I restrict attention to respondents living near anglophone-francophone national borders in West Africa and compare individuals on either side of the same borders, controlling for distance to the border. Second, and most importantly, I turn to Cameroon, which was partitioned between Britain and France after the First World War and reunified at independence, so that the former colonial boundary today runs \textit{within} a single country. Following \citet{Lee_Schultz(2012)} and \citet{Dupraz2019}, I exploit this internal boundary as a geographic discontinuity and compare respondents who live close to either side of it. This design holds constant national institutions, and the boundary itself was drawn hastily by European negotiators, cutting across existing ethnic and religious lines rather than following them \citep[pp.~372]{Lee_Schultz(2012)}.

Three results emerge. First, the anglophone trust premium survives the move from the full cross-country sample to the border samples: near the West African borders, respondents on the anglophone side are 13 to 20 percentage points more likely to trust traditional leaders, and near the internal Cameroonian boundary the gap is 22 to 26 percentage points. Second, the gap is not explained by individual characteristics, ethnolinguistic fractionalization, or distance to the coast and to the capital city, factors that previous work has shown to matter for trust \citep{Ferrara, Nunn_Wantchekon_(2011)}. Third, the attitudinal gap is mirrored by a behavioral one: anglophone respondents are more than twice as likely to report having contacted a traditional leader during the previous year (51 percent against 19 percent in the full sample), which suggests that traditional institutions on the formerly British side are not merely trusted in the abstract but actively used.

I interpret these findings with care. The border comparison identifies the effect of a bundle of characteristics that changed at the partition line, including colonial administrative practice, language of instruction, legal tradition, and the post-independence political position of the anglophone regions; it cannot isolate any single ingredient of that bundle. Section \ref{sect_threats} discusses the main threats to identification, including selective migration, the coarseness of the geographic comparison, and the timing of round 7 of the survey, which was fielded in Cameroon after the beginning of the Anglophone crisis in late 2016. With these caveats stated, the pattern of evidence, in particular the joint behavior of trust and contact, is most consistent with the historical mechanism emphasized in section \ref{mechanism}: indirect rule preserved the authority and the day-to-day governance role of chiefs on the British side, while French administrative practice, by converting chiefs into agents of taxation and labor recruitment, attached a durable stigma to customary authority.

This paper contributes to the literature on the long-run effects of colonial institutions \citep{Hall_Jones_(1999), Acemoglu_al_(2001), Glaeser_Shleifer_(2002), Iyer(2010), Dell(2010), MichalopoulosPapaioannou2013} and, most directly, to the research program that uses the Cameroonian partition as a natural experiment \citep{Lee_Schultz(2012), Dupraz2019}. Relative to \citet{Lee_Schultz(2012)}, who study wealth and local public goods, and \citet{Merima_al_(2018)}, who study national versus ethnic identification, the contribution here is to examine the citizen-chief relationship itself, in both its attitudinal (trust) and behavioral (contact) dimensions, and to connect it to the historiography of native administration.

The remainder of the paper proceeds as follows. Section 2 reviews the related literature. Section \ref{sect_hist_back} provides the historical background on the two forms of colonial rule and on the partition of Cameroon. Section \ref{sect_data} describes the data. Section \ref{sect_emp_approach} presents the empirical approach and the results in three phases: the full cross-country sample (subsection \ref{phase1}), the West African border sample (subsection \ref{phase2}), and the Cameroonian border sample (subsection \ref{phase3}), followed by a discussion of threats to identification (subsection \ref{sect_threats}) and of mechanisms (subsection \ref{mechanism}). Section 6 concludes.

\section{Related Literature}~

This paper sits at the intersection of four literatures: the economics of trust, the persistence of colonial institutions, the comparative study of British and French rule, and the political economy of traditional leadership in contemporary Africa.

\paragraph{Trust and development.} A large body of work establishes that social trust is associated with better economic and political outcomes. \citet{Knack_and_Keefer(1997)} show in a cross-country setting that trust, rather than associational membership, predicts growth; \citet{Zak&Knack(2001)} provide a general equilibrium model and supporting evidence; \citet{Yann_and_Pierre(2010)} identify a causal effect of inherited trust on growth using the trust of descendants of immigrants; and \citet{knack2002}, \citet{uslaner_2002}, and \citet{Christian_2006} document links between trust and the quality of government. At the individual level, \citet{Ferrara} show that traumatic experiences, discrimination, and economic failure depress trust. Closest to the present setting, \citet{Nunn_Wantchekon_(2011)} trace low trust within Africa to the intensity of the slave trades, demonstrating that traumatic historical episodes can shape trust over centuries through internal norms and beliefs. Their outcome measures include trust in relatives, neighbors, and local government councils; trust in traditional leaders, the outcome studied here, is a distinct object, and the slave trade channel they identify is complementary to, not a substitute for, the colonial-rule channel examined here. Within the Afrobarometer Working Papers Series itself, a related line of work examines the determinants of institutional and political trust, including social capital \citep{Kuenzi2008}, perceptions of relative deprivation \citep{Isbell2023}, and media exposure \citep{Bouraima2025}. This paper contributes to this strand by adding formal colonial rule to the list of historical determinants of trust in Africa, and by shifting the object of trust from peers and state institutions to the customary authorities that govern much of daily life on the continent.

\paragraph{Colonial origins and persistence.} A second literature documents that colonial-era institutions cast long shadows. \citet{Hall_Jones_(1999)} and \citet{Acemoglu_al_(2001)} establish the macro-level connection between historically determined institutions and current income. \citet{LaPorta2008} survey the economic consequences of legal origins, and \citet{Glaeser_Shleifer_(2002)} provide the underlying theory of common versus civil law. Within-country designs sharpen the point: \citet{Dell(2010)} shows that the Peruvian mining \textit{mita}, a colonial forced labor institution, depressed household consumption and public goods provision centuries later; \citet{Iyer(2010)} finds that directly ruled areas of India have worse public goods than indirectly ruled areas; \citet{LowesMontero2021} show that the rubber concessions of the Congo Free State, which governed through coercion and co-opted chiefs, produced worse development outcomes and, notably, chiefs of lower quality today; and \citet{MichalopoulosPapaioannou2013, MichalopoulosPapaioannou2014} show both that precolonial ethnic institutions predict current development and that national institutions lose explanatory power far from the capital, which underlines the importance of local, rather than national, governance in Africa. Dependency-school authors reached related conclusions about the damage of colonial extraction by a different route \citep{Frank_(1978), Bagchi_(1982)}. Relative to this literature, which measures persistence mainly in incomes, public goods, and education, the contribution of this paper is to trace the persistence of colonial administrative practice in an attitude, trust in customary authority, together with its behavioral counterpart, contact with chiefs, using the same within-country border variation that gives the strongest designs in this strand their credibility.

\paragraph{British versus French rule.} A third literature compares the legacies of the two major colonizers of Africa. \citet{grier1999colonial} finds that former British colonies grew faster than former French ones; \citet{AthowandRobert2002} document differences in trade patterns; \citet{Frankema2012} shows that educational expansion was greater in British Africa, driven largely by missionary activity; and \citet{FirminSellers2000} cautions, from a comparison of Ghana and Côte d'Ivoire, that metropolitan doctrine translated into practice unevenly, so that the British-French contrast must be treated as a difference in degree rather than kind. Two papers use the same discontinuity exploited here. \citet{Lee_Schultz(2012)} compare outcomes across the former colonial boundary within Cameroon and find that rural anglophone areas have higher wealth and better access to piped water; they attribute the difference to "hard" institutional legacies rather than "soft" cultural ones. \citet{Dupraz2019} uses the same partition to study education and finds that the British advantage was period-specific: it emerged under colonial rule and narrowed, and partly reversed, after independence, a finding that warns against reading any single border gap as an immutable colonial legacy. \citet{CogneauMoradi2014} implement the closely related design on the former Togo partition and show that differences in education and religion across the British-French line emerged already during the interwar years. On the identity dimension, \citet{Merima_al_(2018)} find that anglophone citizens identify less in national terms and more in ethnic terms, and present evidence of weaker state capacity but stronger chiefs on the anglophone side. None of these papers examines trust in traditional leaders, the attitude that most directly reflects the citizen-chief relationship that indirect and direct rule treated so differently. That is the gap this paper fills.

\paragraph{Traditional leaders in contemporary Africa.} Finally, a growing literature in political economy studies chiefs as contemporary political actors rather than colonial relics. \citet{Logan2013} documents, using Afrobarometer data, that popular support for traditional leaders is broad and coexists with support for elected government. \citet{Baldwin2016} argues that chiefs can serve as effective development intermediaries precisely because their long time horizons tie them to their communities. Others are more skeptical: \citet{Mamdani1996} argues that indirect rule created "decentralized despotism," fusing executive, judicial, and administrative power in chiefs who were accountable upward to the colonial state rather than downward to their subjects; \citet{AcemogluReedRobinson2014} show that in Sierra Leone, where fewer ruling families compete for chieftaincies, development outcomes are worse even though respect for authority is higher; and \citet{Ntsebeza2005} documents how unaccountable chieftaincy can obstruct democratization. This debate matters for interpretation: high trust in chiefs is not automatically evidence of good local institutions, a point I return to in section \ref{mechanism}. \citet{Gennaioli2007} show that precolonial political centralization, which conditioned how indirect rule operated, predicts modern public goods provision, and \citet{McNamee2019} finds that indirect rule raised the salience of ethnicity. This paper speaks to this strand by supplying evidence on where the popular standing of chiefs comes from: the cross-sectional surveys of \citet{Logan2013} document that support for traditional leaders is widespread but cannot say why it varies, whereas the border design used here ties part of that variation to a historical cause, colonial administrative strategy, and thereby suggests that chiefly legitimacy today is inherited as well as earned through current performance. On the design side, the paper follows the border-discontinuity tradition of \citet{Dell(2010)}, \citet{Lee_Schultz(2012)}, and \citet{CogneauMoradi2014}; \citet{KeeleTitiunik2015} discuss the specific assumptions required when geographic boundaries serve as sources of identification.

\section{Historical Background\label{sect_hist_back}}

Between the 1870s and 1900, most of Africa was brought under European colonial rule; by the early twentieth century only Ethiopia and Liberia remained independent. Britain and France acquired the largest African empires, and although both ultimately governed through some combination of European officials and African intermediaries, the historiography identifies systematic differences in how they did so. Three differences are most relevant here. First, administrative strategy: British rule relied comparatively more on existing native authorities (indirect rule), while French rule was comparatively more centralized and treated chiefs as subordinate agents of the administration \citep{Crowder1964, Mamdani1996}. Second, labor policy: coerced labor, through the \textit{indig\'enat} disciplinary regime and \textit{prestations} (obligatory labor days), remained a systematic instrument of French administration until its abolition in 1946, whereas British reliance on compulsory labor, while real, was less extensive and was curtailed earlier \citep{Cooper1996}. Third, legal tradition: French colonies inherited codified civil law, while British colonies inherited English common law, with its greater reliance on judicial precedent and, in the colonies, on recognized customary courts \citep{Glaeser_Shleifer_(2002), LaPorta2008}.

Two caveats should be kept in mind throughout. These are tendencies, not hard-and-fast rules: the French also governed through African intermediaries, and the British intervened directly when it suited them \citep{FirminSellers2000}. And doctrine varied across territories and over time within each empire: in settler colonies such as Kenya and Southern Rhodesia, British administration was considerably more direct, while in Senegal the French relationship with the Muslim marabouts operated through accommodation and intermediation in a manner closer to the indirect ideal type. The claim maintained here is only that the average difference in the treatment of customary authority across the two empires was large and systematic, which is what the empirical design requires. The subsections below describe the two styles with these qualifications in place.

\subsection{The British style: indirect rule}

The British applied in Africa their doctrine of indirect rule, which allowed many indigenous institutions to remain in place while introducing British institutions that were often adapted to local custom \citep{Crowder1964}. Administration was comparatively decentralized: British officials relied on native chiefs to implement policy and administer their communities, intervening directly only when circumstances seemed to require it \citep{Crowder1964, grier1999colonial}. As \citet[p.~198]{Crowder1964} explains, ``Though indirect rule reposed primarily on a chief as executive, its aim was not to preserve the institution of chieftaincy as such, but to encourage local self-government through indigenous political institutions, whether these were headed by a single executive authority, or by a council of elders.'' Where no obvious centralized authority existed, the British appointed one: the ``warrant chiefs'' of southeastern Nigeria are the best-known example, and their poor local legitimacy, which erupted in the Aba Women's War of 1929, is a useful reminder that indirect rule worked far better where it could graft onto genuinely rooted institutions than where it invented them. In general, however, the British did not attempt to assimilate Africans to British culture. Colonial policy has been described as an ``economic plan focused on maintaining stability'' \citep{AthowandRobert2002}, under which chiefs and local leaders retained substantial latitude to interpret and apply policy in ways consistent with indigenous standards, and interaction between British officials and the population was limited.

The British likewise left customary law largely in force. ``The very basis of Indirect Rule or the indirect method of British administration by resort to the use of indigenous institutions meant that a total replacement of customary law by English law could not have been contemplated even as a long-term objective'' \citep[pp.~108]{ajayi_1960}. Indigenous tribunals were formally recognized, which allowed customary adjudication to continue throughout the colonial era alongside common law courts. The common law tradition itself provided comparatively stronger protection of individuals and property against the state \citep{Glaeser_Shleifer_(2002), Lee_Schultz(2012)}.

Formal education remained weak and underdeveloped everywhere in colonial Africa, but on average British territories saw greater expansion of primary and secondary schooling for Africans, driven largely by missionary initiative rather than by government policy \citep{Frankema2012, Omolewa}. Many schools taught partly in indigenous languages, and the place of English in the curriculum varied across colonies \citep{Omolewa}. Importantly for the present design, \citet{Dupraz2019} shows for Cameroon that the British educational advantage was concentrated in the colonial period itself and narrowed after independence, so educational legacies should not be presumed permanent.

It must be stressed, following \citet{Mamdani1996}, that preserving chieftaincy is not the same as preserving accountable chieftaincy. By backing chiefs with colonial force while freeing them from many customary checks, indirect rule could concentrate executive, judicial, and administrative power in the same local hands, a configuration Mamdani calls decentralized despotism. Whether the durable standing of chiefs in former British territories reflects earned legitimacy or entrenched authority is an interpretive question that the empirical work below cannot fully resolve, and I return to it in section \ref{mechanism}.

\subsection{The French style: direct rule}

French colonial doctrine, by contrast, aspired first to assimilation: the transformation of colonial subjects into French citizens through the French language, French schooling, and French institutions \citep{grier1999colonial}. In contrast to the structure of indirect rule, the French implemented a comparatively centralized administration \citep{grier1999colonial}. Describing the French colonial experience, \citet{Whittlesey} writes that ``In the administration of the French colonies Frenchmen occupy all the important positions, though properly trained Africans are allowed to fill subordinate posts, and in special circumstances even to become French citizens'' (pp.~363), and that ``The French mode of administration is in theory the flat antithesis of the British. France is in Africa to make Frenchmen out of the Africans. To this end African life is given no official recognition. Administrative officers from France rule directly, native leaders being allowed to handle their own people only by sanction of custom, never of law'' (pp.~367).

In practice, France never had enough European administrators to rule without African intermediaries. The important difference lay in the status of those intermediaries. The French appointed \textit{chefs de canton} and \textit{chefs de village} who derived their authority from the administration rather than from custom, were frequently chosen for loyalty or French education rather than legitimate succession, and were assigned the most resented tasks of colonial government: collecting taxes, recruiting labor, and enforcing the \textit{indig\'enat} \citep{Crowder1964, Cooper1996}. The French undermined the chiefs' autonomy, ``treating them as petty bureaucrats who can be hired and fired at will'' \citep[p.~375]{Lee_Schultz(2012)}. When full assimilation proved too expensive to pursue at scale, policy shifted to the \textit{politique d'association}, but the subordinate position of customary authority was retained. The judicial system was structured after that of metropolitan France, and the civil law tradition emphasized the supremacy of the state over the citizen.

Access to French education was narrow and reserved largely for the would-be African elite. Schooling followed the metropolitan model, was conducted in French from the first year, and left little or no place for indigenous languages. Graduates of this system often became \textit{chefs de canton} or \textit{chefs de quartier} and were in some cases eligible for French citizenship \citep{Crowder1964, Lee_Schultz(2012)}. The overall result was a colonial state that was stricter and more uniform than its British counterpart in the design of administrative, judicial, and educational institutions, and, most relevantly for this paper, one in which customary authority was systematically subordinated and associated with coercion.

\subsection{From colonial practice to contemporary trust\label{sect_theory}}

Why would administrative choices made a century ago still shape trust in traditional leaders today? Each of the three differences described above maps onto a distinct channel with an observable implication.

The first channel runs through the preserved governance functions of chiefs. Under indirect rule, chiefs continued to allocate land, adjudicate disputes, and represent their communities, and these functions largely survived independence in former British territories, where customary authority had an institutionalized place in local administration. Where chiefs do useful things for people, people interact with them, and repeated beneficial interaction is the ordinary foundation of trust in an institution. Under direct rule, by contrast, chiefs were stripped of autonomous functions, so the post-colonial chieftaincy inherited a thinner portfolio and a weaker habit of citizen engagement. This channel predicts that the trust gap should be accompanied by a gap in actual contact with traditional leaders, which is what the data show.

The second channel runs through the association of chiefs with colonial coercion. On the French side, chiefs were the visible enforcers of the most resented colonial policies: they collected taxes, delivered forced labor quotas, and applied the \textit{indig\'enat}. A chief who hands his people to the labor draft damages the standing not only of himself but of the institution he embodies, and attitudes of this kind are transmitted across generations through family and community socialization, in the same way that \citet{Nunn_Wantchekon_(2011)} show slave-trade-era insecurity became a durable norm of mistrust. This channel predicts lower trust on the francophone side even where chieftaincy institutions themselves survived, which is important because the francophone regions in the Cameroonian border sample retain strong chiefdoms.

The third channel runs through legal tradition. Common law practice recognized customary courts, which gave chiefs a formally sanctioned adjudication role that persisted into post-colonial legal systems; the civil law tradition centralized adjudication in state courts \citep{Glaeser_Shleifer_(2002), LaPorta2008}. This channel reinforces the first: it kept chiefs on the British side supplied with a governance function that citizens value and use.

Education and language policy, discussed above because they loom large in the historiography, are not treated here as a fourth channel of their own. Their main relevance is as the leading ``soft'' alternative explanation: if the trust gap merely reflected differences in schooling or Christianization, it should disappear once education and religion are controlled for. The empirical sections therefore control for both, and the survival of the gap under those controls is evidence against the soft channels and in favor of the institutional ones \citep{Lee_Schultz(2012)}.

\section{Data and Preliminary Evidence\label{sect_data}}

\begin{figure}[!htp]
    \centering
    \caption{Countries under British and French rule in the dataset}
    \label{ac}
\includegraphics[width=.7\textwidth]{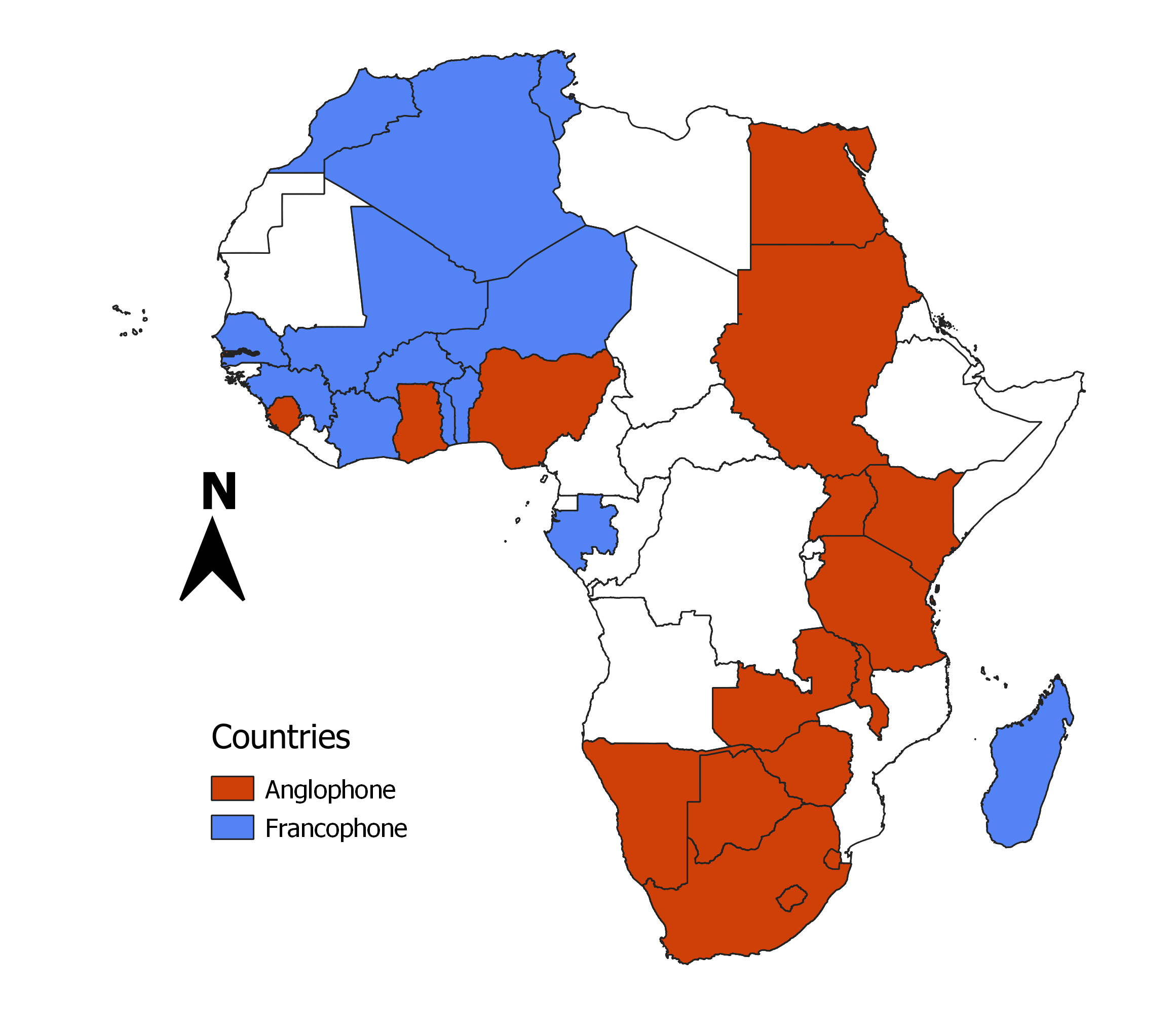}
\end{figure}

The individual-level data used in this study come from Afrobarometer\footnote{Afrobarometer is a non-partisan, pan-African research institution conducting public attitude surveys on democracy, governance, the economy, and society in more than 30 countries on a regular cycle.} survey rounds 6 and 7, covering 31 African countries:\footnote{The 31 countries are: Algeria, Benin, Botswana, Burkina Faso, C\^ote d'Ivoire, Egypt, Gabon, Gambia, Ghana, Guinea, Kenya, Lesotho, Madagascar, Malawi, Mali, Mauritius, Morocco, Namibia, Niger, Nigeria, Senegal, Sierra Leone, South Africa, Sudan, Swaziland, Tanzania, Togo, Tunisia, Uganda, Zambia, and Zimbabwe.} 18 anglophone countries and 13 francophone countries (see Figure \ref{ac}).\footnote{I use ``anglophone'' to refer to a country or region that was colonized by Britain, ``francophone'' for one colonized by France, and, by extension, ``anglophone respondents'' and ``francophone respondents'' for people living in them. The terms are shorthand for colonial heritage, not for language ability: most residents of former British colonies do not speak English and most residents of former French colonies do not speak French. Both terms are written in lower case throughout. The classification is coarse for some cases, notably the North African and settler colonies, whose colonial experience differed substantially from the sub-Saharan pattern described in section \ref{sect_hist_back}; the border analyses below do not rely on these cases. Countries colonized by other powers, notably the former Portuguese and Belgian colonies surveyed by Afrobarometer, are excluded because the paper's comparison is specifically between British and French rule.} I use rounds 6 and 7 because the question on trust in traditional leaders was asked only in these two rounds.

The surveys are based on face-to-face interviews, conducted in local languages, of a national probability sample of 1,200 or 2,400 respondents per round.\footnote{The sampling procedure is described in Appendix \hyperref[appendixA]{A}. There is no reason to be concerned about non-random sampling or response patterns in the surveys.} The same survey rounds were also conducted in Cameroon, which was partitioned between Britain and France; the Cameroonian data are used in subsection \ref{phase3} to address endogeneity concerns. Round 6 was fielded around 2014 and 2015 and round 7 around 2016 to 2018; the implications of this timing for Cameroon are discussed in subsection \ref{sect_threats}.

\begin{table}[!hbtp]
    \centering
      \begin{threeparttable}
    \caption{Overview of the responses to the trust question.\label{trust}}
    \begin{tabular}{ccccccc}
    \hline\hline\\
    & & \multicolumn{5}{c}{ How much do you trust traditional leaders?}\\
   \cline{3-7}    \\
        Response & & \multicolumn{2}{c}{Anglophone} &&\multicolumn{2}{c}{Francophone} \\
         \cline{1-1}\cline{3-4} \cline{6-7}\\
       Not at all &&2,131 &5.98\%&& 3,163 &15.91\% \\
      Just a little &&4,631 &12.99\%&& 3,979 & 20.01\%\\
      Somewhat && 9,003&25.26\%&& 4,436 &22.31\% \\
      A lot && 19,873&55.76\%&& 8,304 &41.77\% \\
      Total &&35,638  &100\%&& 19,882 & 100\%\\
        \hline
    \end{tabular}
      \begin{tablenotes}
      \footnotesize
      \item \textit{Notes:} The table reports the distribution of responses to the trust-in-traditional-leaders question from Afrobarometer survey rounds 6 and 7 (the question is only available in these two rounds). The number of observations is smaller than the number of potential observations because some respondents answered ``don't know'' or are missing. Totals differ slightly from those in Table \ref{tab:summary} because a small number of respondents with valid trust responses lack data on the control variables used in the regressions.
    \end{tablenotes}
      \end{threeparttable}
\end{table}

To measure an individual's trust in traditional leaders, I use a survey question that asks: ``How much do you trust traditional leaders, or haven't you heard enough about them to say?'' Respondents choose among five possible answers: (0) not at all; (1) just a little; (2) somewhat; (3) a lot; and (9) don't know / haven't heard. Table \ref{trust} reports the distribution of responses. The contrast is visible across the entire distribution and is concentrated at the extremes: francophone respondents are nearly three times as likely to answer ``not at all,'' while anglophone respondents are 14 percentage points more likely to answer ``a lot.''

For the remainder of the analysis, I define trust as a binary indicator equal to 1 if the respondent chooses option (2) or (3) and 0 otherwise.\footnote{I obtain similar results when using an alternative index ranging from 0 to 3, where higher values correspond to greater trust.} Respondents who choose (9) are excluded. Table \ref{tab:summary} presents descriptive statistics for trust in traditional leaders and for all other variables used in the analysis. A detailed description of the data sources for each variable is given in Appendix \hyperref[appendixB]{B}.

\begin{table}[!htbp]
    \centering
\small\addtolength{\tabcolsep}{-3pt}
      \begin{threeparttable}
    \caption{Summary statistics.\label{tab:summary}}
    \begin{tabular}{llccccc}
    \hline\hline
         &  &&&  &  &     \\
    &  &\multicolumn{2}{c}{Observations}&&\multicolumn{2}{c}{Mean}\\
   \cline{3-4}\cline{6-7}
   &&Anglo. &Franco.  &&Anglo.  &Franco. \\
   Variables &&(1) &(2)&  &(3)  &(4) \\
         \hline
\multicolumn{2}{l}{\textbf{Main outcome variables: Trust} }    &  &  & &  & \\
      \hspace{.5cm} Traditional leaders && 35,634  & 19,882 && 0.81 (0.39) &0.64 (0.48) \\
\multicolumn{2}{l}{\textbf{Individual controls} }    &&  &   &  &   \\
      \hspace{.5cm} Age && 35,634&19,882&&36.68 (14.86)  & 37.60 (14.56) \\
            \hspace{.5cm} Male &&35,634   &19,882& &0.48 (0.50)& 0.46 (0.50) \\
      \hspace{.5cm} Urban &&35,634    & 19,882 && 0.67 (0.72) &  0.75 (0.73)\\
      \hspace{.5cm} Education &&35,634     & 19,882& & 1.66 (0.93) &  1.17 (1.07)\\
      \hspace{.5cm} Chief contact &&35,560  & 19,865& &0.51 (0.50)& 0.19 (0.39) \\
      \multicolumn{2}{l}{\hspace{.5cm} Living conditions}&35,634  &19,882&&2.65 (1.25)& 2.76 (1.18) \\
    \multicolumn{2}{l}{\textbf{Country controls} }    &&  &   &  &   \\
    \multicolumn{2}{l}{\hspace{.5cm} West Africa}&35,634 &19,882&&0.21 (0.41)& 0.66 (0.47) \\
    \multicolumn{2}{l}{\hspace{.5cm} South Africa}&35,634 &19,882&&0.48 (0.50)& 0.08 (0.28) \\
    \multicolumn{2}{l}{\hspace{.5cm} North Africa}&35,634 & 19,882&&0.04 (0.20)& 0.18 (0.38) \\
        \multicolumn{2}{l}{\hspace{.5cm} East Africa}&35,634 &-&&0.27 (0.45)& - \\
            \multicolumn{2}{l}{\hspace{.5cm} Central Africa}&- &19,882&&-& 0.07 (0.26) \\
        \multicolumn{2}{l}{\hspace{.5cm} Former German colony}&35,634 & 19,882&&0.13 (0.34)& 0.08 (0.27) \\
            \multicolumn{2}{l}{\hspace{.5cm} Landlocked}&35,634 &19,882&&0.41 (0.49)& 0.26 (0.44) \\
   \multicolumn{2}{l}{\textbf{Ethnicity controls} }    &&  &   &  &   \\
         \multicolumn{2}{l}{\hspace{.5cm} Distance to capital (1,000 km)}&35,634  & 19,882& &0.23 (0.19)&0.22 (0.17) \\
      \multicolumn{2}{l}{\hspace{.5cm} Distance to coast (1,000 km)}&35,634  & 19,882&&0.49 (0.35)& 0.30 (0.34) \\
      \multicolumn{2}{l}{\hspace{.5cm} ELF index}&35,634  &19,882&&0.65 (0.23)& 0.62 (0.21) \\
\multicolumn{2}{l}{\textbf{No. of countries} }    &18& 13 && - & -  \\
        \hline
    \end{tabular}
      \begin{tablenotes}
      \footnotesize
      \item \textit{Notes:} This table reports means and standard deviations (in parentheses) of the variables by colonial status (anglophone versus francophone). The control variables are measured at the individual, country, or ethnicity level. ELF\tablefootnote{The $ELF$ index is defined as: $ELF_j = 1 - \sum^I_{i=1}(\frac{n_{ij}}{N_j})^2$, where $n_{ij}$ is the number of people in the $i^{th}$ group in district $j$, $N_j$ is total population in district $j$, and $I$ is the number of ethnolinguistic groups in district $j$.} stands for ethnolinguistic fractionalization, which measures the probability that two randomly selected persons from a given district will not belong to the same ethnolinguistic group. The higher the ELF index, the more fragmented the district. Survey rounds: 6 and 7.
    \end{tablenotes}
      \end{threeparttable}
\end{table}

The first row of Table \ref{tab:summary} shows the means and standard deviations of trust in traditional leaders. Compared with anglophone respondents, the share of francophone respondents who trust traditional leaders is lower by 17 percentage points: 81 percent of anglophone respondents trust traditional leaders, whereas only 64 percent of francophone respondents do so. The table also previews the paper's key mechanism-related fact: 51 percent of anglophone respondents report having contacted a traditional leader during the previous year, against only 19 percent of francophone respondents. Trust in chiefs on the anglophone side is not a purely abstract attitude; it coexists with much more frequent actual engagement with chiefs.

\section{Empirical Approach\label{sect_emp_approach}}

\subsection{Benchmark model\label{phase1}}

I begin by testing whether British colonization is associated with an individual's current level of trust in traditional leaders. The benchmark estimating equation takes the following form:
\begin{equation}\label{model}
    Trust_{ijc} = \alpha + \beta\cdot Anglophone_{c} + X_{ijc}'\Gamma + \delta_{r} + \varepsilon_{ijc}
\end{equation}
where $i$ indexes individuals, $j$ districts, and $c$ countries; $Trust_{ijc}$ is the binary trust indicator defined above; $Anglophone_{c}$ equals 1 for anglophone countries and 0 for francophone countries; and $\delta_{r}$ denotes survey-round fixed effects. The coefficient of interest, $\beta$, captures the difference between anglophone and francophone respondents with respect to the outcome. The vector $X_{ijc}$ collects the individual-, country-, and ethnicity-level controls described in Table \ref{tab:summary}. Because the outcome is binary, equation (\ref{model}) is a linear probability model estimated by OLS, so $\beta$ is directly interpretable as a percentage-point difference; ordered-response models on the original four-point scale are a natural robustness check. The specification is intended to control for variables that might affect trust but are not themselves consequences of colonial status. Gender is obviously not affected by colonial status; cultural variables such as religion and ethnicity may play an important role in shaping willingness to trust, so I control for religious affiliation and for the ethnic composition of the respondent's district. The subscript on $Anglophone_{c}$ makes the identification problem explicit: treatment varies only at the country level, so $\beta$ is identified from cross-country variation among 31 countries rather than from the 55,516 individual observations, country fixed effects cannot be included, and the error term $\varepsilon_{ijc}$ contains a common country-level component, which is why inference should be based on standard errors clustered by country. The benchmark should accordingly be read as descriptive; the border designs below carry the identification burden.

Table \ref{reg} presents the regression results from the benchmark model, using all observations in the sample.
\begin{table}[!htbp]
    \centering
      \begin{threeparttable}
    \caption{Trust and colonial status.\label{reg}}
{
\def\sym#1{\ifmmode^{#1}\else\(^{#1}\)\fi}
\begin{tabular}{l*{4}{c}}
\hline\hline\\
            &\multicolumn{1}{c}{(1)}&\multicolumn{1}{c}{(2)}&\multicolumn{1}{c}{(3)}&\multicolumn{1}{c}{(4)}\\
\hline
Anglophone       &       0.170\sym{***}&      0.0747\sym{***}&       0.111\sym{***}&       0.108\sym{***}\\
            &     [0.004]         &     [0.005]         &     [0.005]         &     [0.005]         \\

 Country controls          &    -        &     Yes       &   Yes       &    Yes              \\
    Individual controls           &    -        &    -              &   Yes      &    Yes             \\
Ethnicity controls           &    -        &   -         &    -        &    Yes             \\
[1em]
Observations       &       55,516         &       55,516         &       55,516         &       55,516         \\
\(R^{2}\)   &       0.036         &       0.093         &       0.122         &       0.126         \\
\hline
\end{tabular}
}
\begin{tablenotes}
      \footnotesize
      \item \textit{Notes:} All regressions include survey-round fixed effects. Heteroskedasticity-robust standard errors are given in brackets; because treatment varies at the country level, these should be regarded as a lower bound, and inference from this table is treated as descriptive (see subsection \ref{sect_threats}). Individual-level controls include age, age squared, a male indicator, an urban indicator, 4 education-level fixed effects, 3 religion fixed effects, 5 living-condition fixed effects, and asset ownership indicators (television, radio, and motor vehicle). The country controls include indicators for region (west, south, central, north, and east Africa), landlockedness, and former German colony status prior to the First World War. The ethnicity controls include distance to the capital city, distance to the coast, and the ELF index. $^{***}$ Significant at the 1 percent level.
\end{tablenotes}
      \end{threeparttable}
\end{table}

The controls capture several variables that could affect individual trust in traditional leaders and whose correlation with colonial status could otherwise confound the results. To capture variation across countries, districts, and individuals, I include controls at the national, district, ethnicity, and individual levels. This flexibility is particularly important in Africa, where state capacity varies considerably across space because central states generally have limited control over areas far from capital cities \citep{MichalopoulosPapaioannou2014}.

In column (1), I report results for equation (\ref{model}) with no controls except survey-round fixed effects. The coefficient on $Anglophone$ is positive and statistically significant: the share of respondents who trust traditional leaders is higher by 17 percentage points in anglophone countries, which restates the mean difference reported in Table \ref{tab:summary}. The logic for each control, and its coding, is described in Appendix \hyperref[appendixB]{B}; in brief, the controls are individual and geographic characteristics that prior work identifies as determinants of trust and that are not themselves consequences of the colonizer's identity. Because ethnic diversity has been shown to depress trust \citep{Ferrara}, I control for ethnolinguistic fractionalization in the respondent's district, constructed from the Afrobarometer sample itself; since Afrobarometer samples are designed to be nationally rather than subnationally representative, this district-level measure is a proxy that is measured with error, and it should be read as such. I also control for proximity to the coast, since distance from the ocean has been shown to matter for trust through historical exposure to the slave trades \citep{Nunn_Wantchekon_(2011)}, and for distance to the capital. When the control sets are added in columns (2) through (4), the estimated coefficient moves but remains positive and statistically significant, between 7 and 11 percentage points. The sensitivity of the point estimate to the control set is itself informative about the potential role of unobservables \citep{Altonji_al(2005), Oster2019}, and is one more reason to prefer the border designs that follow.

\subsection{Border discontinuity analysis for West Africa\label{phase2}}

Despite the inclusion of several controls, the benchmark estimates cannot be read causally, for a reason worth stating plainly. The concern labeled ``endogeneity'' is that colonizer identity was not assigned to territories at random: the places Britain colonized may have differed from the places France colonized before any colonizer arrived, in geography, in precolonial political institutions, or in exposure to earlier historical shocks. For instance, Britain's policy of governing through native administration may have led it to acquire or retain territories that already had strong indigenous political institutions, while France's assimilationist doctrine gave it no such motive. If such selection was operative, current differences between anglophone and francophone countries could predate colonization and would not reflect colonial legacy. No amount of controlling for observable characteristics can rule this out, because the relevant precolonial differences may be unobservable.

\begin{figure}[!htp]
    \centering
    \caption{West African countries for the border discontinuity analysis}    \label{wac}
\includegraphics[width=.7\textwidth]{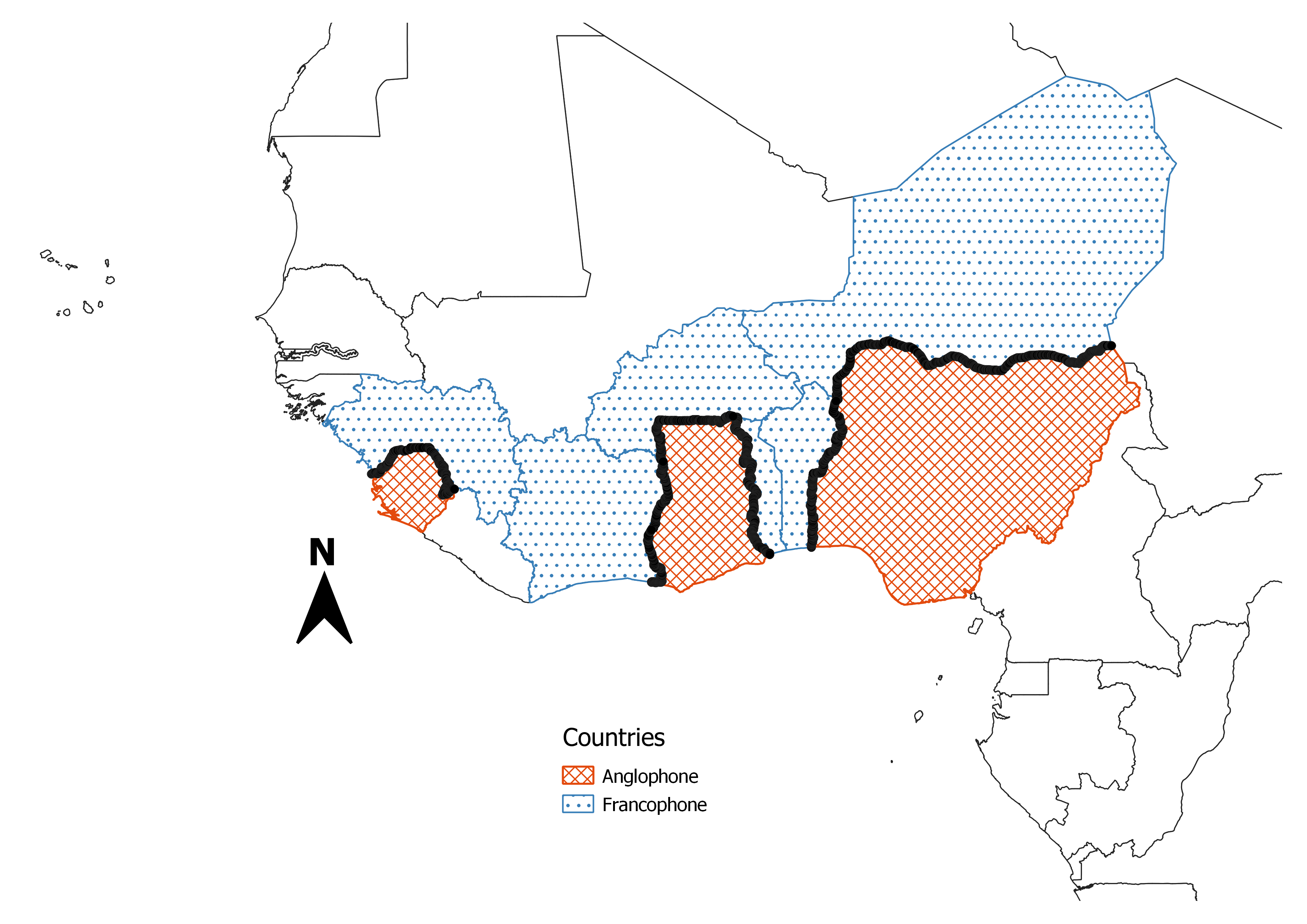}\\
\justify{\footnotesize{\textit{Notes:} The thickest lines show the borders between anglophone and francophone countries. Three of the countries along those borders are anglophone (Ghana, Nigeria, and Sierra Leone) and the rest are francophone (Benin, Burkina Faso, Guinea, Ivory Coast, Niger, and Togo).}}
\end{figure}

The discontinuity approach addresses this problem with a different kind of comparison. The idea is intuitive: people living just on either side of an anglophone-francophone border share essentially the same geography, climate, ecology, and, because colonial borders routinely cut through ethnic homelands, often the same precolonial culture and institutions. What changes discretely at the border is which colonial power ruled. If the location of the border itself was unrelated to local conditions, then comparing respondents who live close to opposite sides of the same border isolates the effect of the colonial assignment from the confounding factors that contaminate cross-country comparisons \citep{KeeleTitiunik2015}. The premise that African borders were drawn with little regard to conditions on the ground is well documented: the partition of the continent was negotiated in Europe, frequently along astronomical lines and hastily surveyed features, and the resulting boundaries split hundreds of ethnic groups \citep{Asiwaju1985, McCauleyPosner2015}. This as-if-random quality is precisely what previous studies exploit at these same borders \citep{Lee_Schultz(2012), CogneauMoradi2014, Dupraz2019}. As a first step, I restrict the sample to respondents who reside near national borders between anglophone and francophone countries and compare individuals across those borders, controlling for distance to the border. I consider the West African countries that share at least one anglophone-francophone border (see Figure \ref{wac}; there are exactly four such borders).

\begin{figure}[!htp]
    \centering
    \caption{Trust in traditional leaders by distance to the border: West Africa}
    \label{RDWestPlot}
    \includegraphics[width=.8\textwidth]{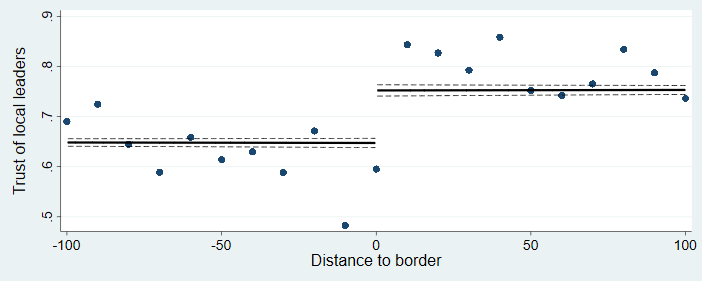}\\
    \justify{\footnotesize{\textit{Notes:} The figure shows, by distance (in km) to the border, the share of respondents who trust traditional leaders. Negative (positive) values represent distance from the border into francophone (anglophone) territory. The distance increases as we move away from 0 on the x-axis. The confidence interval is 95\%.}}
\end{figure}

The estimating equation for the border samples adds the running variable to equation (\ref{model}):
\begin{equation}\label{model_border}
    Trust_{ib} = \alpha + \beta\cdot Anglophone_{i} + f(dist_{ib}) + X_{i}'\Gamma + \delta_{r} + \varepsilon_{ib}
\end{equation}
where $dist_{ib}$ is the distance from respondent $i$'s location to the nearest point of anglophone-francophone border $b$, and $f(\cdot)$ is specified as linear in the estimates reported below. The coefficient $\beta$ now compares respondents on either side of the same boundary, conditional on how far from it they live. Two refinements bring this specification closer to a canonical geographic regression discontinuity design \citep{KeeleTitiunik2015}: allowing the slope of $f(\cdot)$ to differ on the two sides of the border by interacting $dist_{ib}$ with $Anglophone_{i}$, as the fitted lines in Figure \ref{RDWestPlot} already do, and including border-segment fixed effects so that respondents are compared only across the segment nearest to them. The estimates are not sensitive to the distance cutoff or bin width, but implementing these refinements, together with data-driven bandwidth selection \citep{CCT2014}, is a stated priority in subsection \ref{sect_threats}.

A word on the choice of bandwidth. The baseline sample includes respondents within 100 kilometres of a border, and the results are robust to cutoffs of 60, 80, and 120 kilometres. A band of this width is admittedly generous: individuals living 100 kilometres apart on opposite sides of a border are not neighbors, and the wider the band, the weaker the claim that the two groups differ only in colonial assignment. The constraint is statistical power. Afrobarometer enumeration areas are sparse near some border segments, and narrowing the band shrinks the sample quickly; in the Cameroonian analysis below, the border sample already contains only 454 observations. The honest reading is that wider bands buy precision at the cost of comparability, which is why the stability of the estimates as the band narrows from 120 to 60 kilometres matters more than any single specification, and why data-driven bandwidth selection is flagged as a priority for future work.

Two limitations of this design should be stated up front. First, a national border separates not only colonial histories but entire post-independence states: language policy, currency zones, legal systems, and sixty years of national politics all change at the same line, so the estimated gap reflects this full bundle rather than colonial rule alone. Second, one of the francophone border countries, Togo, was a German colony before the First World War, so its colonial history is layered in the same way as Cameroon's \citep{CogneauMoradi2014}; the former-German-colony indicator absorbs part of this, but the West African estimates should be read as descriptive of the anglophone-francophone contrast rather than as clean estimates of a colonial treatment effect. The within-Cameroon analysis of subsection \ref{phase3} addresses the first limitation directly.

The estimates based on the restricted border sample are reported in Table \ref{rdd1}.

\begin{table}[!htbp]
    \centering
      \begin{threeparttable}
    \caption{Trust in traditional leaders: border discontinuity results for West Africa\label{rdd1}}
{
\def\sym#1{\ifmmode^{#1}\else\(^{#1}\)\fi}
\begin{tabular}{l*{5}{c}}
\hline\hline\\
            &\multicolumn{1}{c}{(1)}&\multicolumn{1}{c}{(2)}&\multicolumn{1}{c}{(3)}&\multicolumn{1}{c}{(4)}&\multicolumn{1}{c}{(5)}\\
\hline
Anglophone       &      0.0271\sym{***}&       0.126\sym{***}&       0.128\sym{***}&       0.155\sym{***}&       0.196\sym{***}\\
            &     [0.006]         &     [0.015]         &     [0.015]         &     [0.015]         &     [0.015]         \\
[1em]
Distance to border&-&-&Yes&Yes&Yes\\
Individual controls&-&-&-&Yes&Yes\\
Ethnicity controls&-&-&-&-&Yes\\
[1em]
Observations   &       20,532         &       16,900         &       16,900         &       16,900         &       16,900         \\
\(R^{2}\)   &       0.003         &       0.056         &       0.056         &       0.094         &       0.110         \\
\hline
\end{tabular}
}
      \begin{tablenotes}
      \footnotesize
      \item \textit{Notes:} Column (1) includes all respondents in West Africa. Columns (2) through (5) concern only respondents in the West African border sample, defined as respondents living within 100 kilometres of an anglophone-francophone border. Distance to border is the distance to the nearest anglophone-francophone border. $Anglophone$ is an indicator for whether the observation is from a West African anglophone country. All regressions include survey-round fixed effects. Heteroskedasticity-robust standard errors are given in brackets. The other controls are the same as in Table \ref{reg}. $^{***}$ Significant at the 1 percent level.
    \end{tablenotes}
      \end{threeparttable}
\end{table}

In column (1), I report results with no controls using all francophone and anglophone observations from West Africa, to provide a transparent comparison of the raw mean difference in the region. Columns (2) through (5) use only the observations in the border sample, which includes 16,900 respondents. Column (2) includes no controls; the controls listed in the table are added progressively in the subsequent columns. All estimated coefficients are positive and statistically significant, confirming that the level of trust in traditional leaders is higher among anglophone respondents even within narrow bands around the borders. As in the benchmark, the point estimate grows as controls are added, which again signals that observable differences work against the raw comparison and motivates caution about unobservables \citep{Oster2019}.

The corresponding plot is shown in Figure \ref{RDWestPlot}: the share of respondents who trust traditional leaders is visibly higher on the anglophone side of the borders.

The fitted lines represent the relationship between distance to the national anglophone-francophone borders and trust in traditional leaders, along with 95\% confidence intervals from an OLS regression of trust on distance, estimated separately on each side. Each dot represents the local average (in 10-kilometre bins)\footnote{The results are not sensitive to reasonable changes of the distance cutoff (e.g., 60, 80, or 120 kilometres) or of the bin width (e.g., 5, 15, or 20 kilometres).} of the share of respondents who trust traditional leaders. The advantage of the plot is that it provides a transparent characterization of the data. The pattern is consistent with the estimates in Table \ref{rdd1}: on the anglophone side of the borders, to the right of the centre of the x-axis, the level of trust is higher.

\subsection{The case of Cameroon\label{phase3}}
\begin{figure}[!htp]
    \centering
    \caption{Administrative map of Cameroon}
    \label{map_cameroon}
    \includegraphics[width=.7\textwidth]{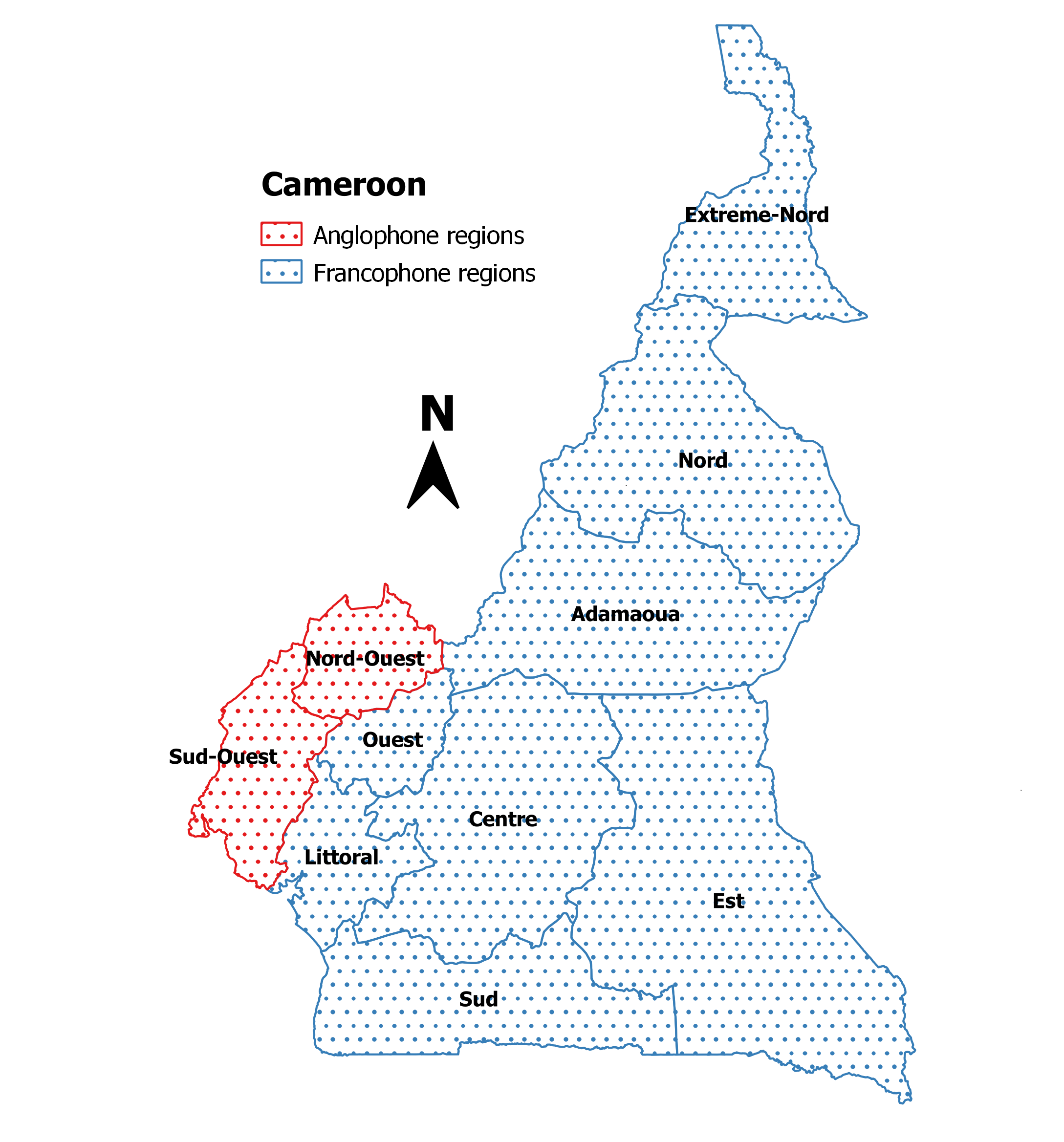}\\
    \centering
    \footnotesize{\textit{Note:} The names on the map are the names of the administrative regions.}
\end{figure}

\begin{figure}[!htp]
    \centering
    \caption{Cameroon, 1901--1972}
    \label{Kamerun}
    \includegraphics[width=.9\textwidth]{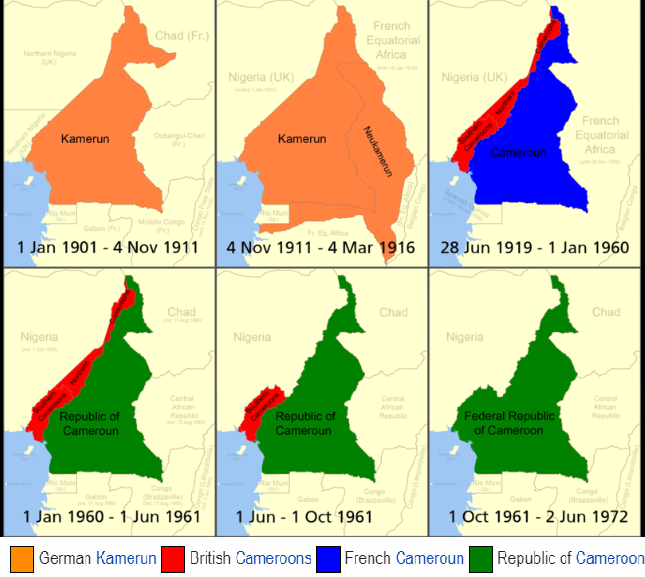}\\
    \centering
   { \footnotesize{\textit{Note:} This figure shows the territorial evolution of Cameroon from 1901 to 1972: the German colony of Kamerun, the 1916--1919 partition into British and French zones, the mandate and trusteeship periods, and reunification after the 1961 plebiscites. Source: adapted from Wikimedia Commons.}}
\end{figure}

Cameroon was colonized by Germany from the mid-1880s. Germany had acquired its empire largely for prestige, and its initial policy emphasized exploration and plantation development over territorial administration \citep[pp.~2--10]{chiabi1997making}. During the First World War, British and French forces captured the German colony and partitioned it between themselves in 1916; the partition was formalized in 1919 and the two zones became League of Nations mandates in 1922 (United Nations trust territories after 1946). France received the larger, eastern share of the territory, while Britain received two narrow, non-contiguous strips along the Nigerian border, the Northern and Southern Cameroons, which it administered as part of Nigeria. In United Nations plebiscites held in February 1961, the Northern Cameroons voted to join Nigeria, while the Southern Cameroons voted to join the newly independent Republic of Cameroon. The two parts formed a federation in October 1961, which was replaced by a unitary state in 1972 (see Figure \ref{Kamerun}). The former Southern Cameroons constitute today's Northwest and Southwest regions, the two anglophone regions among the country's ten administrative regions (see Figure \ref{map_cameroon}).

The Cameroonian case offers a useful setting for discontinuity analysis, for three reasons. First, the variation in colonial rule comes from within Cameroon, which holds country-level differences constant: respondents on both sides of the line live under the same national government, currency, and constitution (see Figure \ref{map_cameroon}).

\begin{figure}[!htp]
    \centering
    \caption{Anglophone and francophone regions sharing the border used in the analysis}
    \label{map}
    \includegraphics[width=.7\textwidth]{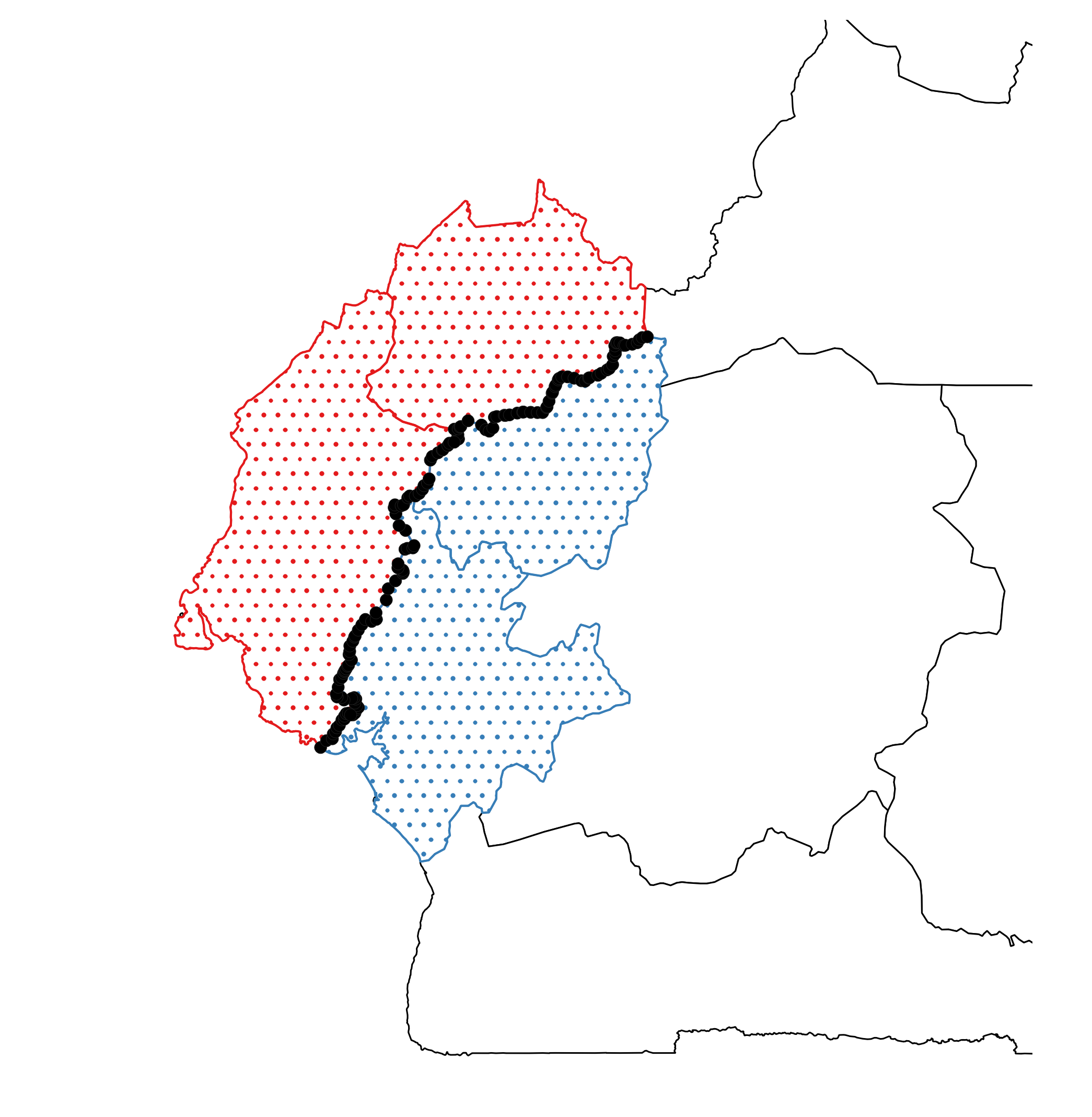}\\
    \centering
\justify{\footnotesize{\textit{Notes:} The colored areas represent the regions from which observations for the border analysis were drawn. The thickest lines show the border between anglophone and francophone regions.}}
\end{figure}

Second, the colonial boundary separating the anglophone and francophone parts was drawn arbitrarily with respect to local conditions (see Figure \ref{map}). Like most colonial borders in Africa, it resulted from hastily arranged agreements between European powers; ``the most notable feature of the colonial border was the degree to which it cut across existing ethnic and religious boundaries'' \citep[pp.~372]{Lee_Schultz(2012)}. In the area used for the analysis, the boundary runs through the culturally connected Grassfields zone, whose societies on both sides share strong and similar chieftaincy traditions, which supports the comparability of the two sides at the moment of partition.

Third, the British and French approaches to colonial rule in Cameroon mirrored the broader patterns described in section \ref{sect_hist_back}. The British administered their zone as an extension of Nigeria and instituted native administration. Comparing the two systems in Cameroon, \citet{chiabi1997making} notes that the ``British had to determine who chiefs were and which areas constituted their jurisdictions. (\ldots) Meanwhile, the French took a different approach. Assessment reports to restructure the country along the lines of chiefdoms was not necessary.''

I undertake the discontinuity analysis, again using equation (\ref{model_border}), on the set of respondents who reside near the administrative border between the anglophone and francophone regions (see Figure \ref{map}), with $dist_{ib}$ now measured to the former partition line within Cameroon. I consider only the francophone regions that share this border with the anglophone regions, namely the Ouest and Littoral regions.\footnote{Although the Adamaoua region touches the former inter-colonial border at its far north, there is no observation near the border in that region, so it is excluded from the analysis.} The key identifying assumption is that, prior to the partition, the areas on the two sides of the line were not systematically different in observable and unobservable factors affecting trust in traditional leaders. The common German colonial baseline supports this assumption, though imperfectly: German administration was itself geographically uneven, concentrated on the coast and in the plantation zone around Mount Cameroon, so the assumption is more plausible for the inland Grassfields section of the boundary used here than for the coastal section \citep{Dupraz2019}. The estimates for Cameroon are reported in Table \ref{rdd2}.

\begin{table}[!htbp]
    \centering
      \begin{threeparttable}
    \caption{Trust in traditional leaders: Cameroon border results\label{rdd2}}
{
\def\sym#1{\ifmmode^{#1}\else\(^{#1}\)\fi}
\begin{tabular}{l*{5}{c}}
\hline\hline\\
            &\multicolumn{1}{c}{(1)}&\multicolumn{1}{c}{(2)}&\multicolumn{1}{c}{(3)}&\multicolumn{1}{c}{(4)}&\multicolumn{1}{c}{(5)}\\
\hline
Anglophone    &       0.150\sym{***}&       0.252\sym{***}&       0.244\sym{***}&       0.225\sym{***}&       0.264\sym{***}\\
            &     [0.028]         &     [0.046]         &     [0.049]         &     [0.062]         &     [0.073]         \\
[1em]
Distance to border&-&-&Yes&Yes&Yes\\
Individual controls&-&-&-&Yes&Yes\\
Ethnicity controls&-&-&-&-&Yes\\
[1em]
Observations      &        2,331         &         454         &         454         &         454         &         454         \\
\(R^{2}\)   &       0.013         &       0.062         &       0.063         &       0.108         &       0.113         \\
\hline
\end{tabular}
}
      \begin{tablenotes}
      \footnotesize
      \item \textit{Notes:} Column (1) includes all respondents in Cameroon from the 10 regions. Columns (2) through (5) concern only respondents in the border sample (Figure \ref{map}). Distance to border is the distance to the nearest point of the former anglophone-francophone boundary within Cameroon. All regressions include survey-round fixed effects. Heteroskedasticity-robust standard errors are given in brackets. The other controls are the same as in Table \ref{reg}. $^{***}$ Significant at the 1 percent level.
    \end{tablenotes}
      \end{threeparttable}
\end{table}

\begin{figure}[!htp]
    \centering
    \caption{Trust in traditional leaders by distance to the border: Cameroon}
    \label{RDCamPlot}
    \includegraphics[width=.8\textwidth]{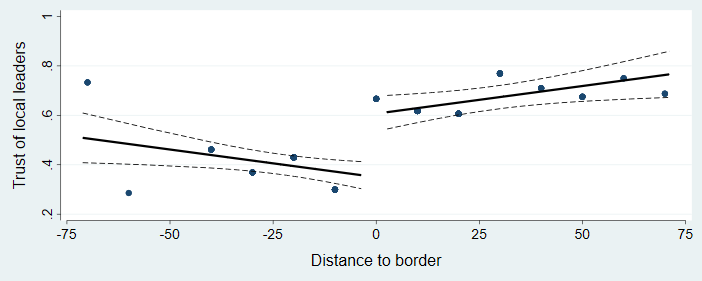}\\
    \justify{\footnotesize{\textit{Notes:} The figure shows, by distance (in km) to the border, the share of respondents who trust traditional leaders. Negative (positive) values represent distance from the border into francophone (anglophone) territory. The distance increases as we move away from 0 on the x-axis. The confidence interval is 95\%.}}
\end{figure}

The results from the Cameroonian sample in Table \ref{rdd2} are similar to those from the West African sample: trust in traditional leaders is significantly higher among anglophone respondents, by 22 to 26 percentage points in the border sample. The magnitude is large but not implausible against the raw national difference of 15 percentage points in column (1) and the descriptive patterns in Table \ref{trust}. The pattern is confirmed by the plot for the Cameroonian sample in Figure \ref{RDCamPlot}, where the fitted lines represent the relationship between distance to the former boundary and trust in traditional leaders with 95\% confidence intervals, and the dots represent local averages in 10-kilometre bins.

\subsection{Threats to identification and interpretation\label{sect_threats}}

Several threats to identification deserve explicit discussion; they define the boundaries of what the estimates can and cannot support.

\paragraph{Specification and inference.} The border estimates in Tables \ref{rdd1} and \ref{rdd2} come from OLS on distance-restricted samples with a linear distance control, which is the transparent specification pioneered for this setting by \citet{Lee_Schultz(2012)}, rather than from a full nonparametric regression discontinuity estimator with data-driven bandwidths \citep{CCT2014, KeeleTitiunik2015}. The stability of the estimates across distance cutoffs and bin widths is reassuring, but a complete treatment would add local linear estimation with separate slopes on each side, optimal bandwidth selection, tests for covariate smoothness and for bunching of respondents at the boundary \citep{McCrary2008}, and border-segment fixed effects in the West African sample so that respondents are compared only across the same border. On inference, the reported standard errors are heteroskedasticity-robust; because treatment varies at the country level in the benchmark and at the region level in Cameroon, clustered and spatially dependent errors would be more conservative, and the benchmark results in particular should be read as descriptive rather than as precisely estimated causal effects. The multi-level structure of the data, with a country-level explanatory variable and individual-level outcomes, could alternatively be modeled directly with a hierarchical (multilevel) linear model with country random effects, which would deliver appropriately conservative country-level inference by construction; this is a natural companion specification for the benchmark.

\paragraph{Compound treatment.} Even within Cameroon, the former colonial boundary bundles several treatments: colonial administrative practice, but also language of schooling, legal tradition, and, since 1961, the distinct political position of the anglophone regions within the Cameroonian state. The estimates identify the effect of falling on the British side of the partition line, understood as this entire bundle; disentangling its ingredients requires additional variation, such as the timing evidence exploited by \citet{Dupraz2019}.

\paragraph{Migration and sorting.} Sixty years of internal migration mean that current residence near the boundary is potentially endogenous. If individuals with particular attitudes toward traditional authority sorted across the line, the border contrast would partly reflect selection. The Afrobarometer data locate respondents at their current residence; restricting attention to respondents interviewed in their region of origin, or assigning treatment by ethnic homeland as in \citet{Nunn_Wantchekon_(2011)}, are the natural robustness checks and are a priority for future drafts.

\paragraph{Timing and the Anglophone crisis.} Round 7 of the Afrobarometer was fielded in Cameroon in 2018, after the beginning of the Anglophone crisis in late 2016, a period of protest and then armed conflict concentrated precisely in the Northwest and Southwest regions. The crisis plausibly raised the salience of anglophone identity and of traditional institutions, and it complicated survey operations in the affected areas. A contemporaneous political shock of this kind could inflate the measured trust gap for reasons unrelated to colonial legacy. The essential robustness check is to estimate the Cameroonian discontinuity separately for round 6, which was fielded before the crisis; a gap already present in round 6 cannot be attributed to the conflict. I flag this check as a requirement for any causal reading of the Cameroonian results.

\paragraph{Interpretation of the outcome.} Finally, higher trust in traditional leaders should not be equated automatically with better local institutions. As \citet{Mamdani1996} and \citet{AcemogluReedRobinson2014} emphasize, chiefly authority preserved by indirect rule could be unaccountable, and deference is not the same as institutional quality. The estimates in this paper measure the citizen-chief relationship; welfare conclusions require additional evidence.

\subsection{Mechanisms and alternative explanations\label{mechanism}}
As described in section \ref{sect_hist_back}, the literature on colonial legacies suggests two families of channels: ``hard'' institutional legacies, operating through the structure of native administration and labor policy, and ``soft'' legacies, operating through religion, education, and culture \citep{Lee_Schultz(2012)}.

The hard-legacy interpretation runs as follows. British indirect rule preserved, and where necessary created, native authorities with real governance functions; chiefs on the British side continued to allocate land, resolve disputes, and represent their communities, and the traditional order retained its normative claims on the population. On the French side, by contrast, customary authority was subordinated to the administration, and chiefs were tasked with collecting taxes and recruiting forced labor under the \textit{indig\'enat} and the \textit{prestations} system until 1946 \citep{Cooper1996, Crowder1964}. This assignment mattered for legitimacy: a chief who delivers his people to the colonial labor draft is a chief whose standing suffers, even where, as in the Bamil\'ek\'e chiefdoms of the Ouest region, the chieftaincy institution itself survived colonial rule intact. The francophone side of the Cameroonian border sample is, notably, a zone of strong surviving chiefdoms; what differs across the line is not the existence of chiefs but their historical relationship to coercion and, plausibly for that reason, the trust they command. On this reading, the legacy operates less through the destruction of institutions than through a durable stain on their legitimacy.

Two pieces of evidence in the data are consistent with this interpretation, though neither is decisive. First, the trust gap is mirrored by a large gap in actual engagement: anglophone respondents are more than twice as likely to have contacted a traditional leader in the previous year (Table \ref{tab:summary}), and this contact gap is visible near the borders as well (Figures \ref{grdd_west} and \ref{grdd_cam} in the appendix). Trust travels with use, as one would expect if chiefs on the anglophone side retained real governance functions. Second, the gap survives controls for education, religion, and living conditions, which weighs against the soft-legacy channels operating through schooling and Christianization; this echoes the conclusion of \citet{Lee_Schultz(2012)} for material outcomes. A formal analysis of the contact outcome, and a decomposition of the trust gap by cohort to test whether it is fading or stable, are natural next steps; \citet{Dupraz2019} shows for education that colonial-era gaps at this border need not persist unchanged, so persistence itself is something to establish rather than assume.

Alternative explanations remain. The post-independence marginalization of the anglophone regions may have pushed citizens toward traditional institutions as a counterweight to a distant central state; the legal-origins channel \citep{LaPorta2008} may operate through the recognized status of customary courts on the common law side; and the crisis-timing concern of subsection \ref{sect_threats} applies with full force. What the evidence does allow one to say is that the difference is not explained by observable individual characteristics, that it appears within a single country at a boundary drawn with no regard for local conditions, and that it involves behavior as well as attitudes.

\section{Conclusion}
This paper provides evidence that the form of European colonial rule left a mark on the relationship between African citizens and their traditional leaders. Respondents in former British colonies are substantially more likely to trust traditional leaders than respondents in former French colonies. The gap survives the move from cross-country comparisons to comparisons across anglophone-francophone borders in West Africa, and it appears, with a magnitude of 22 to 26 percentage points, at the former colonial partition line within Cameroon, where national institutions are held constant and the boundary was drawn without regard to local conditions. The attitudinal gap is accompanied by a behavioral one: anglophone respondents contact traditional leaders far more often.

The literature on colonialism and African history suggests why the legacy of British rule, compared with French rule, may support trust in traditional leaders. Britain governed through a decentralized system that preserved the governance functions of chiefs, while French administration subordinated customary authority to the central state and assigned chiefs the enforcement of its most unpopular policies, taxation and forced labor above all. On this interpretation, the French legacy operates through a durable association between chieftaincy and coercion rather than through the disappearance of chiefs, an interpretation consistent with the survival of strong chiefdoms on the francophone side of the Cameroonian border.

The limits of the evidence should be stated as plainly as the findings. The border design identifies the effect of a bundle of characteristics that changed at the partition line; the estimates rest on modest samples with simple specifications; and the timing of the most recent survey round, which overlaps the Anglophone crisis, requires that the Cameroonian results be verified on pre-crisis data before a causal reading is secure. Establishing the round-by-round robustness of the discontinuity, implementing a full nonparametric border design with appropriate inference, and testing the mechanism through the behavior of the contact outcome are the priorities for future work. Whether the trust that chiefs command on the anglophone side reflects accountable local institutions or entrenched authority, a question posed sharply by \citet{Mamdani1996} and \citet{AcemogluReedRobinson2014}, remains open, and answering it would connect this historical legacy to its consequences for development today.

\bibliographystyle{apalike}
\bibliography{mybib_revised}

@book{chiabi1997making,
  title={The Making of Modern Cameroon},
  author={Chiabi, E.},
  lccn={97033470},
  series={The Making of Modern Cameroon},
  url={https://books.google.com/books?id=7tXcAQAACAAJ},
  year={1997},
  publisher={University Press of America}
}

@book {Putnam2000,
	title = {Bowling Alone: The Collapse and Revival of American Community},
	year = {2000},
	note = {Translations into Swedish (Stockholm: SNS, 2001); Spanish (Barcelona: Galaxia Gutenberg, 2002); Italian (Bologna: Il Mulino, 2004); Japanese (Tokyo: Kashiwashobo, 2006); Chinese (Peking University Press, 2006); Estonian (Tallinn: Hermes, 2007); Polish (Warsaw: Wydawnictwa Akademickie I Profesjonalne, 2008); Serbian (Novi Sad: Mediterran, 2008), Korean (Seoul: Paperroad Publishers, 2008).},
	publisher = {Simon \& Schuster},
	organization = {Simon \& Schuster},
	address = {New York},
	url = {http://bowlingalone.com/},
	author = {Robert D. Putnam}
}

@article {Ferrara,
	title = {Who Trusts Others?},
	journal = {Journal of Public Economics},
	volume = {85},
	year = {2002},
	pages = {207-34},
	author = {Alberto Alesina and Eliana La Ferrara}
}

@article{Omolewa,
author = {Omolewa, Michael},
title = {Educating the "Native": A Study of the Education Adaptation Strategy in British Colonial Africa, 1910-1936},
journal = {The Journal of African American History},
volume = {91},
number = {3},
pages = {267-287},
year = {2006},
doi = {10.1086/JAAHv91n3p267},
URL = {https://doi.org/10.1086/JAAHv91n3p267},
eprint = {https://doi.org/10.1086/JAAHv91n3p267}
}

@article{Whittlesey,
 ISSN = {00157120},
 URL = {http://www.jstor.org/stable/20028773},
 author = {Derwent Whittlesey},
 journal = {Foreign Affairs},
 number = {2},
 pages = {362--373},
 publisher = {Council on Foreign Relations},
 title = {British and French Colonial Technique in West Africa},
 volume = {15},
 year = {1937}
}

@article{ajayi_1960, 
title={The Interaction of English Law with Customary Law in Western Nigeria: II}, 
volume={4}, 
DOI={10.1017/S0021855300002692}, 
number={2}, 
journal={Journal of African Law}, 
publisher={Cambridge University Press}, 
author={Ajayi, Jacob F. Ade}, 
year={1960}, 
pages={98–114}
}

@article{AthowandRobert2002,
 ISSN = {87553449},
 URL = {http://www.jstor.org/stable/45194064},
 author = {Brian Athow and Robert G. Blanton},
 journal = {Journal of Third World Studies},
 number = {2},
 pages = {219--241},
 publisher = {University Press of Florida},
 title = {COLONIAL STYLE AND COLONIAL LEGACIES: TRADE PATTERNS IN BRITISH AND FRENCH AFRICA},
 volume = {19},
 year = {2002}
}

@article{grier1999colonial,
  title={Colonial legacies and economic growth},
  author={Grier, Robin M},
  journal={Public choice},
  volume={98},
  number={3-4},
  pages={317--335},
  year={1999},
  publisher={Springer}
}

@article{Crowder1964,
 ISSN = {00019720, 17500184},
 URL = {http://www.jstor.org/stable/1158021},
 author = {Michael Crowder},
 journal = {Africa: Journal of the International African Institute},
 number = {3},
 pages = {197--205},
 publisher = {[Cambridge University Press, International African Institute]},
 title = {Indirect Rule: French and British Style},
 volume = {34},
 year = {1964}
}

@article{knack2002,
 ISSN = {00925853, 15405907},
 URL = {http://www.jstor.org/stable/3088433},
 author = {Stephen Knack},
 journal = {American Journal of Political Science},
 number = {4},
 pages = {772--785},
 publisher = {[Midwest Political Science Association, Wiley]},
 title = {Social Capital and the Quality of Government: Evidence from the States},
 volume = {46},
 year = {2002}
}

@book{uslaner_2002, 
place={Cambridge}, 
title={The Moral Foundations of Trust}, 
DOI={10.1017/CBO9780511614934}, 
publisher={Cambridge University Press}, 
author={Uslaner, Eric M.}, 
year={2002}
}

@article{Christian_2006,
title = {The multiple facets of social capital},
journal = {European Journal of Political Economy},
volume = {22},
number = {1},
pages = {22 - 40},
issn = {0176-2680},
doi = {https://doi.org/10.1016/j.ejpoleco.2005.05.006},
url = {http://www.sciencedirect.com/science/article/pii/S0176268005000509},
author = {Christian Bj{\o}rnskov},
year = {2006}
}

@book{Mamdani1996,
    author = {Mahmood Mamdani},
    title = {Citizen and Subject: Contemporary Africa and the Legacy of Late Colonialism},
    publisher = {Princeton University Press},
    address = {Princeton},
    year = {1996}
}

@article{Dupraz2019,
    author = {Dupraz, Yannick},
    title = {French and {British} Colonial Legacies in Education: Evidence from the Partition of {Cameroon}},
    journal = {The Journal of Economic History},
    volume = {79},
    number = {3},
    pages = {628--668},
    year = {2019},
    doi = {10.1017/S0022050719000299}
}

@article{CogneauMoradi2014,
    author = {Cogneau, Denis and Moradi, Alexander},
    title = {Borders That Divide: Education and Religion in {Ghana} and {Togo} Since Colonial Times},
    journal = {The Journal of Economic History},
    volume = {74},
    number = {3},
    pages = {694--729},
    year = {2014},
    doi = {10.1017/S0022050714000576}
}

@article{AcemogluReedRobinson2014,
    author = {Acemoglu, Daron and Reed, Tristan and Robinson, James A.},
    title = {Chiefs: Economic Development and Elite Control of Civil Society in {Sierra Leone}},
    journal = {Journal of Political Economy},
    volume = {122},
    number = {2},
    pages = {319--368},
    year = {2014},
    doi = {10.1086/674988}
}

@article{Logan2013,
    author = {Logan, Carolyn},
    title = {The Roots of Resilience: Exploring Popular Support for {African} Traditional Authorities},
    journal = {African Affairs},
    volume = {112},
    number = {448},
    pages = {353--376},
    year = {2013},
    doi = {10.1093/afraf/adt025}
}

@book{Baldwin2016,
    author = {Kate Baldwin},
    title = {The Paradox of Traditional Chiefs in Democratic Africa},
    publisher = {Cambridge University Press},
    address = {New York},
    year = {2016}
}

@article{MichalopoulosPapaioannou2013,
    author = {Michalopoulos, Stelios and Papaioannou, Elias},
    title = {Pre-colonial Ethnic Institutions and Contemporary {African} Development},
    journal = {Econometrica},
    volume = {81},
    number = {1},
    pages = {113--152},
    year = {2013},
    doi = {10.3982/ECTA9613}
}

@article{MichalopoulosPapaioannou2014,
    author = {Michalopoulos, Stelios and Papaioannou, Elias},
    title = {National Institutions and Subnational Development in {Africa}},
    journal = {The Quarterly Journal of Economics},
    volume = {129},
    number = {1},
    pages = {151--213},
    year = {2014},
    doi = {10.1093/qje/qjt029}
}

@article{FirminSellers2000,
    author = {Firmin-Sellers, Kathryn},
    title = {Institutions, Context, and Outcomes: Explaining {French} and {British} Rule in {West Africa}},
    journal = {Comparative Politics},
    volume = {32},
    number = {3},
    pages = {253--272},
    year = {2000},
    doi = {10.2307/422366}
}

@book{Asiwaju1985,
    editor = {A. I. Asiwaju},
    title = {Partitioned {Africans}: Ethnic Relations Across {Africa}'s International Boundaries, 1884--1984},
    publisher = {C. Hurst},
    address = {London},
    year = {1985}
}

@article{LowesMontero2021,
    author = {Lowes, Sara and Montero, Eduardo},
    title = {Concessions, Violence, and Indirect Rule: Evidence from the {Congo Free State}},
    journal = {The Quarterly Journal of Economics},
    volume = {136},
    number = {4},
    pages = {2047--2091},
    year = {2021},
    doi = {10.1093/qje/qjab021}
}

@article{McNamee2019,
    author = {McNamee, Lachlan},
    title = {Indirect Colonial Rule and the Salience of Ethnicity},
    journal = {World Development},
    volume = {122},
    pages = {142--156},
    year = {2019},
    doi = {10.1016/j.worlddev.2019.05.019}
}

@article{LaPorta2008,
    author = {La Porta, Rafael and Lopez-de-Silanes, Florencio and Shleifer, Andrei},
    title = {The Economic Consequences of Legal Origins},
    journal = {Journal of Economic Literature},
    volume = {46},
    number = {2},
    pages = {285--332},
    year = {2008},
    doi = {10.1257/jel.46.2.285}
}

@article{Frankema2012,
    author = {Frankema, Ewout H. P.},
    title = {The Origins of Formal Education in Sub-{Saharan Africa}: Was {British} Rule More Benign?},
    journal = {European Review of Economic History},
    volume = {16},
    number = {4},
    pages = {335--355},
    year = {2012},
    doi = {10.1093/ereh/hes009}
}

@book{Cooper1996,
    author = {Frederick Cooper},
    title = {Decolonization and {African} Society: The Labor Question in {French} and {British} {Africa}},
    publisher = {Cambridge University Press},
    address = {Cambridge},
    year = {1996}
}

@article{CCT2014,
    author = {Calonico, Sebastian and Cattaneo, Matias D. and Titiunik, Roc\'io},
    title = {Robust Nonparametric Confidence Intervals for Regression-Discontinuity Designs},
    journal = {Econometrica},
    volume = {82},
    number = {6},
    pages = {2295--2326},
    year = {2014},
    doi = {10.3982/ECTA11757}
}

@article{KeeleTitiunik2015,
    author = {Keele, Luke J. and Titiunik, Roc\'io},
    title = {Geographic Boundaries as Regression Discontinuities},
    journal = {Political Analysis},
    volume = {23},
    number = {1},
    pages = {127--155},
    year = {2015},
    doi = {10.1093/pan/mpu014}
}

@article{McCrary2008,
    author = {McCrary, Justin},
    title = {Manipulation of the Running Variable in the Regression Discontinuity Design: A Density Test},
    journal = {Journal of Econometrics},
    volume = {142},
    number = {2},
    pages = {698--714},
    year = {2008},
    doi = {10.1016/j.jeconom.2007.05.005}
}

@article{Oster2019,
    author = {Oster, Emily},
    title = {Unobservable Selection and Coefficient Stability: Theory and Evidence},
    journal = {Journal of Business \& Economic Statistics},
    volume = {37},
    number = {2},
    pages = {187--204},
    year = {2019},
    doi = {10.1080/07350015.2016.1227711}
}

@book{Ntsebeza2005,
    author = {Lungisile Ntsebeza},
    title = {Democracy Compromised: Chiefs and the Politics of the Land in {South Africa}},
    publisher = {Brill},
    address = {Leiden},
    year = {2005}
}

@article{Gennaioli2007,
    author = {Gennaioli, Nicola and Rainer, Ilia},
    title = {The Modern Impact of Precolonial Centralization in {Africa}},
    journal = {Journal of Economic Growth},
    volume = {12},
    number = {3},
    pages = {185--234},
    year = {2007},
    doi = {10.1007/s10887-007-9017-z}
}
\section*{Appendix}

\subsection*{A. Sampling procedures\label{appendixA}}
This section summarizes the sampling procedures used by Afrobarometer. A full description of these procedures is available on their website at \href{https://www.afrobarometer.org/surveys-and-methods/sampling-principles}{afrobarometer.org/surveys-and-methods/sampling-principles}.

Afrobarometer uses national probability samples designed to generate a representative cross-section of all citizens of voting age in a given country. First, random selection methods are used at every stage of sampling:
\begin{itemize}
    \item Draw secondary sampling units (SSUs).
\item Randomly select primary sampling units (PSUs).
\item Randomly select sampling start points.
\item Interviewers randomly select households.
\item Within the household, the interviewer randomly selects an individual respondent. Each interviewer alternates between interviewing a man and interviewing a woman to ensure gender balance in the sample.
\end{itemize}
Second, sampling is conducted with probability proportionate to population size wherever possible, to ensure that more populated geographic units have a proportionally greater probability of being chosen into the sample.

The randomly selected sample of 1,200 (2,400) people gives a margin of error of 2.8 (2)\% at a 95\% confidence level.

\subsection*{B. Variables\label{appendixB}}

This appendix describes each variable, its coding, and the logic for its inclusion. The controls fall into three groups. Individual controls capture personal characteristics that prior work identifies as determinants of trust \citep{Ferrara} and that are not consequences of colonizer identity. Country controls absorb broad geographic and historical differences across states. Ethnicity controls capture features of the respondent's local and historical environment, included because ethnic diversity and slave-trade exposure have been shown to depress trust \citep{Ferrara, Nunn_Wantchekon_(2011)}.

\textit{Anglophone:} An indicator for whether (or not) the observation is from an anglophone region or country.

\textit{Age:} The respondent's age in years; age squared is included to allow a nonlinear profile.

\textit{Male:} An indicator equal to 1 if the respondent is male.

\textit{Urban:} An indicator for whether the respondent's enumeration area is classified as urban by Afrobarometer.

\textit{Education:} The respondent's highest education level, collapsed into four categories (no formal schooling, primary, secondary, post-secondary), entered as fixed effects.

\textit{Religion:} The respondent's religious affiliation, collapsed into three groups entered as fixed effects.

\textit{Living conditions:} The respondent's self-assessed present living conditions on the standard five-point Afrobarometer scale, from very bad to very good, entered as fixed effects.

\textit{Asset ownership:} Indicators for household ownership of a television, a radio, and a motor vehicle.

\textit{Distance to coast:} Smallest distance between each household and the nearest coast.\footnote{Author's calculation using QGIS software. \label{ff}}

\textit{Distance to the border:} Smallest distance between each household and the closest anglophone-francophone border.\textsuperscript{\ref{ff}}

\textit{Distance to capital city:} Smallest distance between each household and the capital city.\textsuperscript{\ref{ff}} Data on capital cities are from Geodatos.\footnote{\href{https://www.geodatos.net}{https://www.geodatos.net.}}

\textit{Landlocked:} A binary indicator equal to 1 if the country is landlocked.

\textit{Former German colony:} A dummy variable for Tanzania, Namibia, and Togo, which were German colonies prior to the First World War.

\textit{Chief contact:} A binary indicator equal to 1 if the respondent reports having contacted a traditional leader about some important problem or to give them their views during the past year.

\textit{ELF index:} The ethnolinguistic fractionalization index, which measures the probability that two randomly selected persons from a given district will not belong to the same ethnolinguistic group. It is defined as: $ELF_j = 1 - \sum^I_{i=1}(\frac{n_{ij}}{N_j})^2$, where $n_{ij}$ is the number of people in the $i^{th}$ group in district $j$, $N_j$ is total population in district $j$, and $I$ is the number of ethnolinguistic groups in district $j$. The higher the ELF index, the more fragmented the district. Because the index is constructed from the Afrobarometer sample, which is designed to be nationally rather than subnationally representative, it is a noisy proxy for true district-level diversity and is interpreted as such.

All the remaining variables used in this paper come from Afrobarometer rounds 6 and 7, downloadable from the official Afrobarometer website at \href{https://www.afrobarometer.org/data/merged-data}{afrobarometer.org/data/merged-data}. The subnationally geocoded Afrobarometer data are available on request.

\newpage
\subsection*{C. Figures}

\begin{figure}[!htp]
    \centering
    \caption{Contact with traditional leaders and living conditions by distance to the border: West Africa}
    \label{grdd_west}
    \includegraphics[width=.6\textwidth]{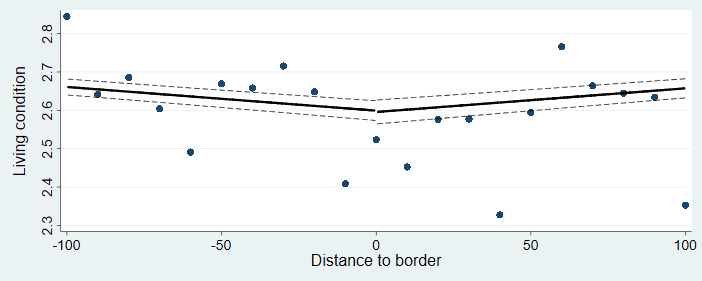}\\
        \includegraphics[width=.6\textwidth]{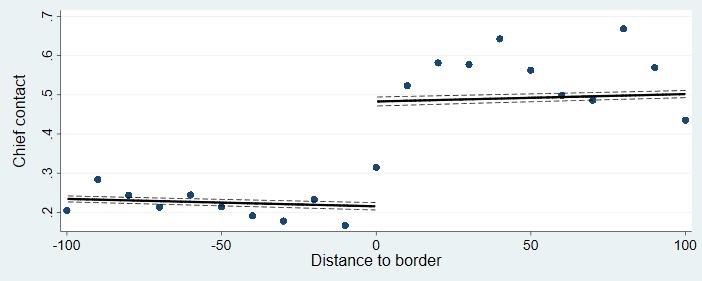}\\
    \footnotesize{\textit{Notes:} Negative (positive) values represent distance from the border into francophone (anglophone) territory. The distance increases as we move away from 0 on the x-axis. The confidence interval is 95\%.}
\end{figure}

\begin{figure}[!htp]
    \centering
    \caption{Contact with traditional leaders and living conditions by distance to the border: Cameroon}
    \label{grdd_cam}
    \includegraphics[width=.6\textwidth]{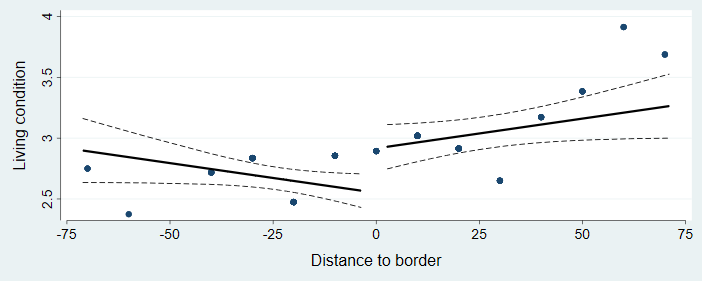}\\
        \includegraphics[width=.6\textwidth]{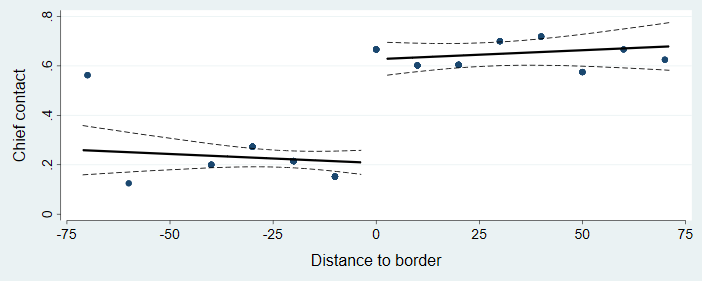}\\
    \footnotesize{\textit{Notes:} Negative (positive) values represent distance from the border into francophone (anglophone) territory. The distance increases as we move away from 0 on the x-axis. The confidence interval is 95\%.}
\end{figure}

\end{document}